\documentclass[twocolumn,prb,superscriptaddress,aps,showpacs]{revtex4-1}
\usepackage{graphicx} 
\usepackage{longtable} 
\usepackage{natmove}
\begin{document}

\title{High-Throughput Screening of Perovskite Alloys for\\Piezoelectric Performance and Formability}

\author{R.~Armiento} \affiliation{Department of Physics, Chemistry and Biology (IFM), Link{\"o}ping University, SE-58183 Link{\"o}ping, Sweden.}

\author{B.~Kozinsky} \affiliation{Research and Technology Center, Robert Bosch LLC, Cambridge MA 02142, USA.}

\author{G.~Hautier} \affiliation{Institut de la Mati{\`e}re condens{\'e}e et des Nanosciences (IMCN), European Theoretical Spectroscopy Facility (ETSF), Universit{\'e} Catholique de Louvain, Louvain-la-Neuve, Belgium}

\author{M.~Fornari} \affiliation{Department of Physics, Central Michigan University, Mt. Pleasant, Michigan 48859, USA.}

\author{G.~Ceder} \affiliation{Department of Materials Science and Engineering, Massachusetts Institute of Technology, Cambridge, Massachusetts 02139, USA.}

\date{\today}

\begin{abstract}
We screen a large chemical space of perovskite alloys for systems with the right properties to accommodate a morphotropic phase boundary (MPB) in their composition-temperature phase diagram, a crucial feature for high piezoelectric performance. We start from alloy end-points previously identified in a high-throughput computational search. An interpolation scheme is used to estimate the relative energies between different perovskite distortions for alloy compositions with a minimum of computational effort. Suggested alloys are further screened for thermodynamic stability. The screening identifies alloy systems already known to host a MPB, and suggests a few new ones that may be promising candidates for future experiments. Our method of investigation may be extended to other perovskite systems, e.g., (oxy-)nitrides, and provides a useful methodology for any application of high-throughput screening of isovalent alloy systems. 
\end{abstract}

 \maketitle

\section{Introduction}
Materials with high piezoelectric coefficients are of significant technological interest. The anomalously large piezoelectric effect in lead zirconate titanate (PZT)-based compounds \cite{overview-1,overview-2,overview-3,overview-4,pzt-1,pzt-2,pzt-3} has generated much attention in the literature, and much focus has been on replacing PZT with an environmentally friendly, lead-free, substitute with similar properties \cite{prev1,prev2,prev3,prev4,prev5,prev6,prev7,prev8,prev9,prev10,prev11}. To the authors knowledge, despite major efforts, there exists to date no lead-free technological replacement for PZT. Two notable recent lead-free isovalent alloy systems have been published by Saito {\it et al.}
\cite{toyota},
$(\mathrm{K},\mathrm{Na},\mathrm{Li})(\mathrm{Nb},\mathrm{Ta})\mathrm{O}_{3}$,
and by Liu and Ren \cite{prl},
$\mathrm{Ba}(\mathrm{Ti}_{0.8}\mathrm{Zr}_{0.2})\mathrm{O}_{3}$-$(\mathrm{Ba}_{0.7}\mathrm{Ca}_{0.3})\mathrm{Ti}\mathrm{O}_3$. While these materials are reported to show a similar phase boundary as PZT, which is crucial for a high piezoelectric effect, the former material is expensive and difficult to synthesize due to
problems with vaporization of potassium oxide during sintering
\cite{KNO-problem-1, KNO-problem-2}, and the latter has a phase boundary that is
very temperature dependent. To find new alloy compounds that reproduces the relevant properties of PZT provides a well-defined challenge for the growing community in high-throughput computational material design. \cite{Curtarolo_Morgan_Persson_Rodgers_Ceder_2003, Greeley_Jaramillo_Bonde_Chorkendorff_Norskov_2006, Ortiz_Eriksson_Klintenberg_2009, Gavartin_Sarwar_Papageorgopoulos_Gunn_Garcia_Perlov_Krzystala_Ormsby_Thompsett_Goldbeck-Wood_Andersen_French_2009, Munter_Landis_Abild-Pedersen_Jones_Wang_Bligaard_2009, Norskov_Abild-Pedersen_Studt_Bligaard_2011, Jain_Hautier_Moore_Ping_Ong_Fischer_Mueller_Persson_Ceder_2011, Curtarolo_Setyawan_Hart_Jahnatek_Chepulskii_Taylor_Wang_Xue_Yang_Levy_Mehl_Stokes_Demchenko_Morgan_2012, Curtarolo_Setyawan_Wang_Xue_Yang_Taylor_Nelson_Hart_Sanvito_Buongiorno-Nardelli_Mingo_Levy_2012, Curtarolo_Setyawan_Wang_Xue_Yang_Taylor_Nelson_Hart_Sanvito_Buongiorno-Nardelli_Mingo_Levy_2012, Bennett_Garrity_Rabe_Vanderbilt_2012, Roy_Bennett_Rabe_Vanderbilt_2012, Bennett_Rabe_2012, Curtarolo_Hart_Nardelli_Mingo_Sanvito_Levy_2013, Kirklin_Meredig_Wolverton_2013, Klintenberg_Eriksson_2013, materialsgenomese, aflowliborg, materialsproject}

The morphotropic phase boundary (MPB) is a phase region in the temperature-composition phase diagram where the macroscopic polarization vector quasi-continuously changes direction \cite{polarization-vanderbildt, rotationpolarization, fucohen}. This region is responsible for the high piezoelectric coefficient in PZT which occurs at nearly equal concentrations of $\mathrm{Pb}\mathrm{Zr} \mathrm{O}_3$ (PZ) and $\mathrm{Pb}\mathrm{Ti}\mathrm{O}_3$ (PT) \cite{mpb-1,mpb-2,mpb-monoclinic-1,overview-4,rotationpolarization, fucohen,polarization-vanderbildt}. We recently presented a high-throughput framework for the computation of material properties of pure ternary perovskites, which was used to identify a set of alloy endpoints with beneficial properties for forming an MPB \cite{armientoprb}. The present work expands on the prior identification of alloy end-points by analyzing the possible solid solutions and aiming at a nearly exhaustive computational search for any stable binary isovalent perovskite alloy that can form an MPB. 

\begin{figure}
\begin{center}
\includegraphics[width=246.0pt]{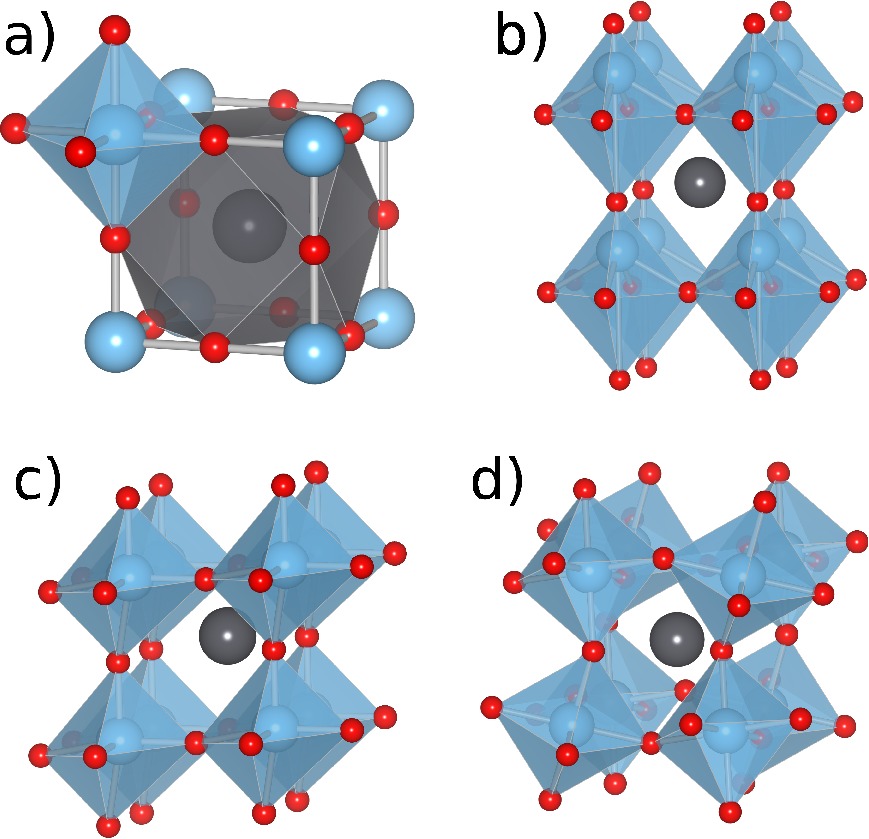}
\caption{(color online) The ideal perovskite crystal structure (a) and the three main distortions (b-d) considered in our prior \cite{armientoprb} and present work. The ideal perovskite structure $\mathrm{ABO}_3$ has a  12-coordinated $A$-site cation (big, black) and a 6-coordinated $B$-site cations (big, pale shaded/blue), embedded in octahedra (blue) defined by oxygens (small, dark/red). The distorted phases are (b) tetragonal (TET), (c) rhombohedral (RHO), and (d) rotations along the (1,1,1) direction (ROT). For illustrative purposes the magnitude of the distortions are exaggerated compared to typical values. The figures were created using \textsc{VESTA} \cite{vesta}.}
\label{fig:perovskite}
\end{center}
\end{figure}

Our investigation focuses on the perovskite crystal structure, since we know, from PZT, that this structure can accommodate the crucial phase-change driven polarization rotation that is coupled to strain. The screening for alloy end-points in our previous work used two primary criteria connected to the appearance of an MPB in an isovalent alloy system. First, each alloy end-point has to be an insulator or semiconductor with a suitably large band gap to avoid current leakage. This criteria was realized by selecting compounds with a quite generous limit on the minimum Kohn-Sham band gap, $\Delta_\mathrm{KS} > 0.25\ \mathrm{eV}$. Furthermore, three main distortions of the ideal perovskite cell was considered: tetragonal (TET), rhombohedral (RHO) and octahedral rotation (ROT) around the (1,1,1) direction. The ideal perovskite crystal structure and the considered distortions are illustrated in Fig.~\ref{fig:perovskite}. The second screening criteria used was a limit on the difference in the lowest and highest formation energy of these distortions, i.e., that
\begin{eqnarray}
\Delta E^\mathrm{P} &=& \max(E_\mathrm{ROT},E_\mathrm{TET},E_\mathrm{RHO}) - \nonumber\\ 
&&\quad\quad\min(E_\mathrm{ROT},E_\mathrm{TET},E_\mathrm{RHO}) < 1.00\ \mathrm{eV}
\end{eqnarray}
The motivation for limiting this energy difference is that it is a necessary prerequisite to obtain a small energy barrier between phases necessary for a MPB (See Ref.~\onlinecite{armientoprb} for an extended discussion). Since the focus in our previous work was on identifying alloy end-points rather than specific useful compositions, no screening on thermodynamic stability was done in that work. 

Our prior screening identified a set of $49$ pure perovskite systems to use as alloy end-points. If we were to freely consider alloys of up to four different such end-points, the number of possible alloys is $49^4 = 5,764,801$. These alloys can then be considered over a large set of relative composition ratios, resulting in a very large configuration space. Our strategy of screening this vast space is the following: we use an interpolation scheme based on computed phase energies for distorted phases of the pure alloy end-points to predict the energetics of a possible combination without any further time-consuming computation. We use the interpolation scheme to identify two-component perovskite compositions that at some mixing ratio prefer a TET distortion (i.e., compositions that in this respect are similar to PT.) The reason for this focus is that previously identified MPBs in perovskites appear in the region between such a TET distorted phase and one of the other distortions. For any combination of two perovskites that is predicted by the interpolation to favor TET distortion in its zero temperature ground state at \emph{any} mixing ratio, we investigate computationally all mixing ratios that are possible to realize in a 40-atom supercell (i.e, concentrations in steps of $1/8$). By comparing computed energies of these systems with those of competing phases we can determine the stability of each realized mixing ratio. A stable TET material provides an opportunity for an MBP through further adjustment of the alloying ratio to reach an equilibrium between distorted phases.

The rest of the paper is structured as follows. In section II we discuss the interpolation scheme used to predict the properties of alloys. In section III we discuss the methods used for investigating competing phases for stability. In Section IV the results are discussed, and relevant alloys are identified. Section V summarizes our main conclusions.

\section{Interpolation of Alloy Properties}\label{sec:interpolation}
Our previous high-throughput screening of alloy end-points has resulted in a large database of energies for the various distorted phases of perovskites at various unit cell volumes, computed with density functional theory (DFT) \cite{dft-1,dft-2}. In past literature DFT has generally been successful in describing the properties of ferro- and piezoelectric perovskite compounds \cite{dftpiezo-1,dftpiezo-2,polarization-vanderbildt,dftpiezo-4,dftpiezo-5,dftpiezo-6,dftpiezo-7,dftpiezo-8,ghita}. In the following we propose a scheme for estimating the phase energy of a distorted phase of any alloy based on these computed energies. 

One may first consider a straightforward linear interpolation between the phase energies of the distortions of the involved alloy end-points. However, these phase energies depend on the cell volume, sometimes in a quite non-linear way. At a volume when a distorted phase becomes equal to the cubic phase (i.e., there is no longer an energy preference to distort the structure), the phase energy vs.\ volume curve of the distorted phase joins that of the cubic phase, giving a discontinuous derivative. One example of the importance of the volume dependence of the phase energies of alloy end-points is the PT-like material $(\mathrm{Ba}_{0.7}\mathrm{Ca}_{0.3})\mathrm{Ti}\mathrm{O}_3$ in the MPB alloy by Liu and Ren \cite{prl}. As pointed out in our previous perovskite screening work \cite{armientoprb} neither $\mathrm{Ba}\mathrm{Ti}\mathrm{O}_3$ or $\mathrm{Ca}\mathrm{Ti}\mathrm{O}_3$ are predicted to prefer a TET distortion at their respective relaxed volumes (cf.,~Fig.~5a of that work). Hence, a linear interpolation of the phase energies would not predict an alloy that prefers a TET distortion in this system. However, if we take into account that the larger cell volume of $\mathrm{Ba}\mathrm{Ti}\mathrm{O}_3$ expands the volume of $\mathrm{Ca}\mathrm{Ti}\mathrm{O}_3$, we find that for some alloy ranges the TET distortion is the lowest energy distortion (cf.~Fig.~5b of Ref.~\onlinecite{armientoprb}). From this example it should be clear that the volume dependence of the distorted phase energies must be explcitly taken into account.

Hence, in the present work we use the following interpolation scheme: consider a specific alloy $X_xY_{1-x}$ between end-points $X$ and $Y$ at mixing ratio $x$. First, we estimate the resulting unit cell volume $V$ by linear interpolation from the cell volumes of $X$ and $Y$ in accordance with Vegard's law \cite{vegardslaw}. Second, for each of the 49 suggested alloy end-points we have a set of 11 computations spanning cell volumes from $45$ to $90\ \textrm{\AA}^3$. We use these pre-computed energies for $X$ and $Y$ to construct piecewise linear energy vs.\ volume curves for the two end-points separately, and use them to obtain the distorted phase energies at volume $V$. Finally, these end-point phase energies at volume $V$ are interpolated linearly to alloy ratio $x$ to estimate the phase energies of the alloy $X_xY_{1-x}$.

One may consider the use of a higher-order interpolation in the energy vs.\ volume curve to improve accuracy. However, as explained above, the curve may have a discontinuous derivative at some points, and thus a piecewise linear interpolation is a safe option. The performance and reliability of this scheme will be discussed as we investigate its results for known systems in Sec.~\ref{sec:Results}.

\section{Stability screening}
The interpolation scheme in the previous section can be used to identify alloys with any combination of energies of distorted phases without having to run computations for each separate alloy and composition ratio. One of the primary requirements to experimentally synthesize an alloy is that its thermodynamical instability is not too large. We base our screening on a zero temperature distance to the hull analysis using the high-throughput phase diagram methods presented in Refs.~\onlinecite{phaseshyue, Jain_Hautier_Moore_Ping_Ong_Fischer_Mueller_Persson_Ceder_2011}. We will use the term `distance to the hull' to refer to the energy difference between a phase and the most stable linear combination of competing phases at the same composition.

A minor thermodynamical instability at zero temperature does not necessarily make synthesis impossible, if, e.g., the energy barriers to competing phases are large enough, or if the entropy contribution sufficiently helps stabilizing the alloy. Hence, rather than outright dismissing compounds with nonzero hull distance, we take the distance to the hull as an estimate of the probability for a specific alloy to be formable. This is intended to take into account the possible errors in the computational methods, metastability, and a reasonable entropy contribution at relevant temperatures. Based on experience with the predictability of these methods, and in particular, based on obtained values for known stable perovskite alloys \cite{geoffroy0}, we will in this work dismiss compounds as extremely unlikely to be formable when they are $0.1\ \mathrm{eV/atom}$ above the hull. For the contribution from error in the computational methods at $0\ \mathrm{K}$, a closer examination for ternary oxides have shown the standard deviation in formation energies of $0.024\ \mathrm{eV/atom}$, meaning that 90\% of the errors should lie with $0.047\ \mathrm{eV/atom}$ \cite{geoffroy0}.

To avoid under-predicting the distance to the hull, formation energies for as many competing phases as possible need to be included. The structure prediction tool of Ref.~\onlinecite{geoffroy2010} was used to predict possible competing phases of each of our alloy end-points. To find reliable convergence settings, pseudopotentials, etc., is an ongoing effort across the high-throughput computational community \cite{Curtarolo_Morgan_Persson_Rodgers_Ceder_2003, Greeley_Jaramillo_Bonde_Chorkendorff_Norskov_2006, Ortiz_Eriksson_Klintenberg_2009, Gavartin_Sarwar_Papageorgopoulos_Gunn_Garcia_Perlov_Krzystala_Ormsby_Thompsett_Goldbeck-Wood_Andersen_French_2009, Munter_Landis_Abild-Pedersen_Jones_Wang_Bligaard_2009, Norskov_Abild-Pedersen_Studt_Bligaard_2011, Jain_Hautier_Moore_Ping_Ong_Fischer_Mueller_Persson_Ceder_2011, Curtarolo_Setyawan_Hart_Jahnatek_Chepulskii_Taylor_Wang_Xue_Yang_Levy_Mehl_Stokes_Demchenko_Morgan_2012, Curtarolo_Setyawan_Wang_Xue_Yang_Taylor_Nelson_Hart_Sanvito_Buongiorno-Nardelli_Mingo_Levy_2012, Curtarolo_Setyawan_Wang_Xue_Yang_Taylor_Nelson_Hart_Sanvito_Buongiorno-Nardelli_Mingo_Levy_2012, Bennett_Garrity_Rabe_Vanderbilt_2012, Roy_Bennett_Rabe_Vanderbilt_2012, Bennett_Rabe_2012, Curtarolo_Hart_Nardelli_Mingo_Sanvito_Levy_2013, Kirklin_Meredig_Wolverton_2013, Klintenberg_Eriksson_2013}. Sepecifically in this work, we calculate the energies of all found competing phases closely following the recipes in Ref.~\onlinecite{Jain_Hautier_Moore_Ping_Ong_Fischer_Mueller_Persson_Ceder_2011}. The structure prediction algorithm is not guaranteed to be absolutely exhaustive, and it is possible that some relevant competing phases have been overlooked. This would mean it is possible for us to predict certain alloys as formable, whereas a synthesis attempt would instead find the omitted competing phase. An omitted competing phase will not lead us to incorrectly dismiss a system.

For each identified perovskite alloy that the interpolation scheme has predicted to prefer TET distortion at some mixing ratio, we perform a DFT computation for all systems in the full range of mixing ratios for the $A$ and $B$ site that can be realized in a 40-atom supercell, i.e., in steps of $1/8$, relaxed starting from a weakly TET distorted crystal (i.e., $c/a = 1.03$). This starting point for the relaxation means the results below for stability vs.\ composition will specifically show the stability of the TET distortion. However, the energy difference between the distortions is usually small compared to that of competing phases (which is expected since our screening for end-points included that the energy difference between the distortions are limited, i.e. $\Delta E^\mathrm{P} < 1.00\ \mathrm{eV}$). 

The ordering of different ions in the DFT computations are chosen in a way where similar ions are placed as far apart as possible in the supercell. Due to the similarity in electrostatics and bonding between the ions in an isovalent alloy, the energy difference of different orderings are expected to be small on the energy scale of thermodynamical stability. However, this assumption is not well tested and may be problematic. Future work is planned to more systematically examine the energy dependence on ordering in isovalent alloys. 

\section{Results}\label{sec:Results}
We select a pair of the 49 alloy end-points that have the same element on either the A or B site, and use the energy interpolation scheme to identify all pairs of alloy end-points for which the TET distortion has the lowest energy of the considered distortions, at some mixing ratio. Since 11 out of our 49 end-points already prefer the TET distortion in their pure phase, they will trivially generate TET alloys with very small amounts of any one of the other end-points. We are unlikely to learn more from these cases in addition to what was already discussed in Ref.~\onlinecite{armientoprb}, and hence we disregard alloys predicted to be TET only up to a very small ratio with another end-point ($< 1\%$). From this procedure 112 entries remain to be investigated for thermodynamical stability as described in the previous section. 

In our analysis below, we dismiss without further investigation any alloy that, over all mixing ratios, exceeds a hull distance of $0.1\ \mathrm{eV}$. We have also found it very difficult to reach convergence of the electron density in computations of $\mathrm{Rb(Nb,Ta)O}_3$ and $\mathrm{(Rb,Cs)NbO}_3$, which has prevented further analysis of those alloy systems. The result is $42$ alloy systems to be investigated. We first investigate compounds known to exhibit an MPB, and then discuss the most interesting alloy systems found in the screening. The complete data set of all 42 alloys satisfying the screening criteria are published alongside this paper as supplemental material \cite{supplemental}. 

\subsection{The PZT system}

\begin{figure}
\begin{center}
\includegraphics[width=246.0pt]{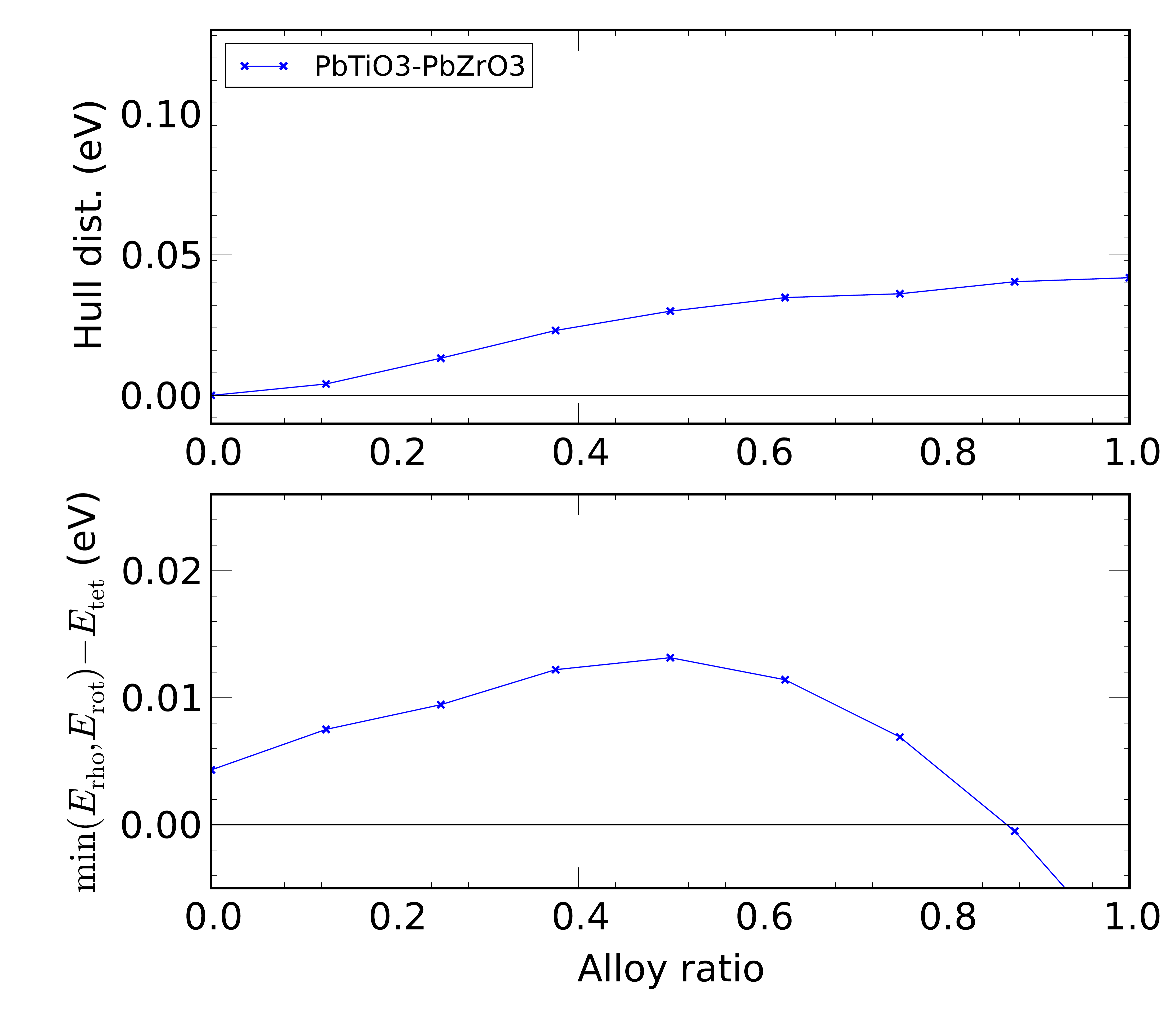}
\caption{(color online) Stability (upper graph) and energy preference for TET distortion (lower graph) in eV/atom for the two component alloy $\mathrm{PbTiO}_3-\mathrm{PbZrO}_3$ (PT-PZ). The upper graph identifies these alloys as possibly stable over the whole range (below $50\ \mathrm{meV}$/atom), and the lower graph shows how the alloys over a large range of relative ratios prefer TET distortion (since the lower graph is $> 0$), pointing to the PZT alloy as an interesting piezoelectric material possible to synthesize (which is known to be true in this case). The comparably low stability of pure PZ and the misplaced crossover between the TET to RHO distortions for PT and PZ are discussed in the text.}
\label{fig:pzt}
\end{center}
\end{figure}

Figure~\ref{fig:pzt} shows the distance to the hull and preference for a TET distortion of $\mathrm{Pb(Ti_xZr_{1-x})O}_3$. This prototypical piezoelectric system serves as a good example of our analysis and provides a test case for the accuracy of the predictions. The upper graph of Fig.~\ref{fig:pzt} shows the distance to the hull stays below $50\ \mathrm{meV}$/atom over all alloy ratios, but it also appears PZ is predicted as significantly less stable than PT. The lower graph in the figure shows a preference for the TET distortion for mixing ratios $(1-x) < 0.84$.

The results corroborate that the accuracy of the interpolation scheme is sufficient to qualitatively identify PZT as having a crossover point between TET and RHO distortion depending on the ratio of PZ to PT. However, the quantitative accuracy is limited; the crossover point is here predicted at $(1-x) \approx 0.840$ rather than $0.5$ as known for PZT. Furthermore, while PT is predicted to have a zero hull distance, i.e., being thermodynamically stable, we find a hull distance for pure PZ of $\sim 40\ \mathrm{meV/atom}$. Some of this hull distance can be accounted for due to our deliberate choice of computing the hull distance for a TET distorted phase. We find the most stable phase of PZ in the materials project database \cite{materialsproject} at the time of writing to be at $\sim 30\ \mathrm{meV}$/atom above the hull. This is still a large hull distance for a system known to be stable, but is not unreasonable when compared to previous accuracy tests of these methods \cite{geoffroy0}. The hull distance for PZ further motivates the choice in this work of considering combinations of perovskite end points up to as much as $100\ \mathrm{meV/atom}$ above the hull at zero temperature before outright dismissing them as thermodynamically unstable.

\subsection{The BTZC system}
\begin{figure}
\begin{center}
\includegraphics[width=246.0pt]{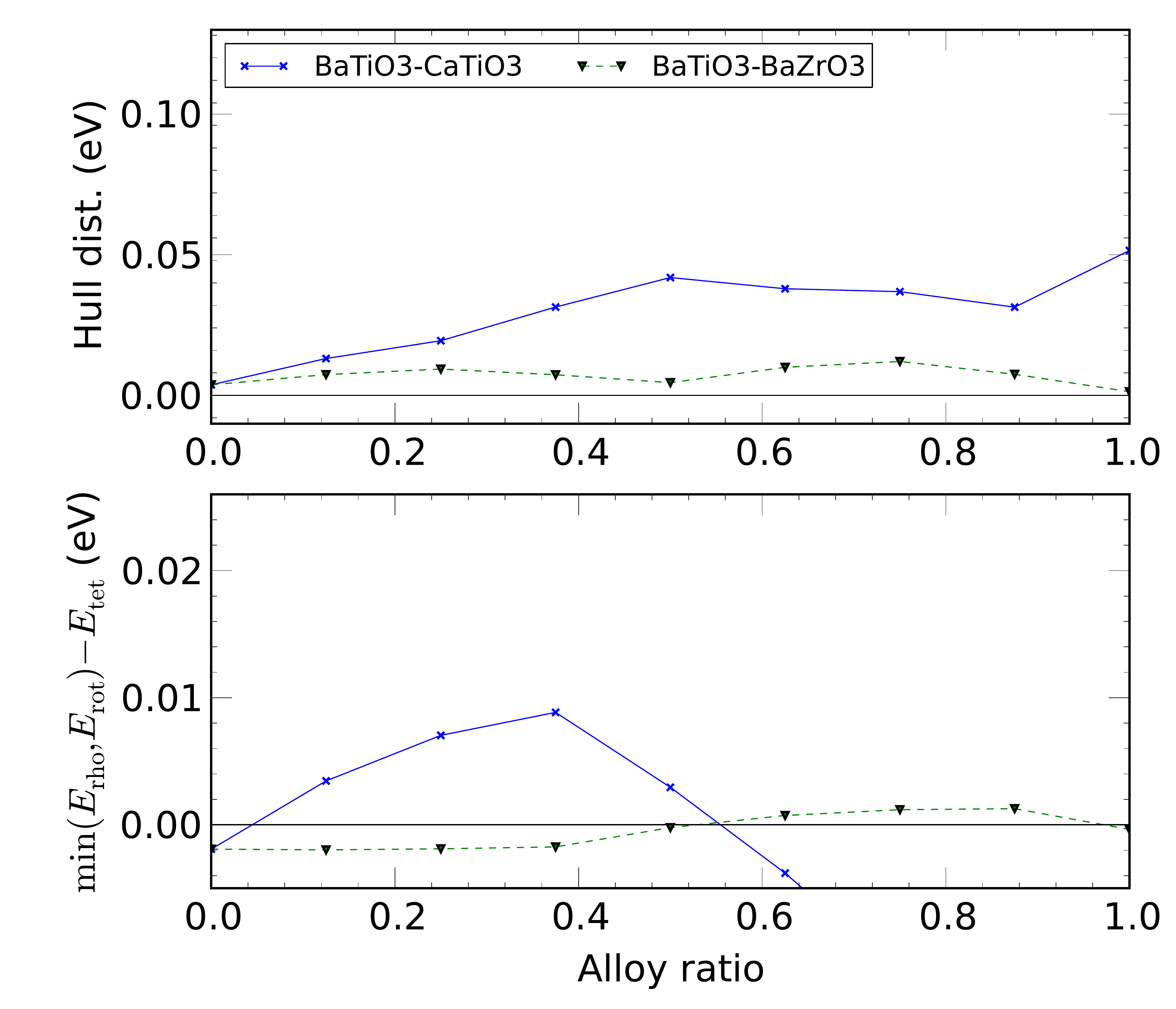}
\caption{(color online) Stability (upper graph) and energy preference for TET distortion (lower graph)  in eV/atom for two-component alloys related to the $\mathrm{Ba}\mathrm{Ti}\mathrm{O}_{3}-\mathrm{Ba}\mathrm{Zr}\mathrm{O}_{3}-\mathrm{Ca}\mathrm{Ti}\mathrm{O}_3$ system. The diagram points at $\mathrm{Ba}\mathrm{Ti}\mathrm{O}_{3}-\mathrm{Ca}\mathrm{Ti}\mathrm{O}_3$ being possibly stable, and preferring TET distortion, as was reported by Liu and Ren \cite{prl}.}
\label{fig:prl}
\end{center}
\end{figure}

As discussed in the introduction of this work, Liu and Ren have identified an MPB in the alloy system of $\mathrm{Ba}\mathrm{Ti}\mathrm{O}_{3}-\mathrm{Ba}\mathrm{Zr}\mathrm{O}_{3}-\mathrm{Ca}\mathrm{Ti}\mathrm{O}_3$ \cite{prl}. Figure~\ref{fig:prl} shows our stability and distorted phase results for perovskite alloys involved in this material. The lower graph of Fig.~\ref{fig:prl} shows that $(\mathrm{Ba}_x\mathrm{Ca}_{1-x})\mathrm{Ti}\mathrm{O}_3$ is predicted to prefer a TET distortion for mixing ratios $(1-x)<0.58$, and for those ratios, the hull distance of the TET distorted phase (shown in the upper graph) stays below $50\ \mathrm{meV/atom}$. The second perovskite alloy shown in Fig.~\ref{fig:prl}, $\mathrm{Ba}(\mathrm{Ti}_x\mathrm{Zr}_{1-x})\mathrm{O}_3$ has a quite small hull distance across all alloy ratios (upper graph), weakly prefers the TET distortion at mixing ratio $(1-x)>0.5$, and otherwise weakly prefers a non-TET distortion (seen in the lower graph).

These results corroborates the work of Liu and Ren. We can view their material as a 4-component solid solution created out of two 2-component end-points. This work focuses on the alloy end-point with TET distortion (i.e., the PT-like material), which in the work of Liu and Ren is $(\mathrm{Ba}_x\mathrm{Ca}_{1-x})\mathrm{Ti}\mathrm{O}_3$ with mixing ratio $(1-x) = 0.3$. Our conclusion from Fig.~\ref{fig:prl} is that this is a thermodynamically stable alloy that prefers the TET distortion. For the opposite alloy end-point (the PZ-like material) Liu and Ren used $\mathrm{Ba}(\mathrm{Ti}_x\mathrm{Zr}_{1-x})\mathrm{O}_3$ with mixing ratio $(1-x) = 0.2$. Our results suggests this alloy to be stable and have a weak non-TET distortion, and thus an alloy between these two end-points appear to be a good candidate for a MPB.

While discussing the details of the interpolation scheme above, we mentioned that $(\mathrm{Ba}_x\mathrm{Ca}_{1-x})\mathrm{Ti}\mathrm{O}_{3}$ prefers the TET distortion at certain mixing ratios $x$, despite none of the end-points preferring a TET distortion. We note that this behavior is precisely reproduced by our interpolation scheme, as seen in the lower part of Fig.~\ref{fig:prl} (and the range for alloys preferring TET distortion is predicted as $0.02 < (1-x) < 0.58$.) It is a validation of the interpolation scheme that this qualitative behavior is reproduced. We also note that for the $\mathrm{Ca}\mathrm{Ti}\mathrm{O}_{3}$ end-point, the ground state is a very distorted monoclinic perovskite. The energy difference between this and the TET distortion is quite large and accounts for the instability seen in the hull distance graph for this end-point.

\subsection{The KNLNT system}

\begin{figure}
\begin{center}
\includegraphics[width=246.0pt]{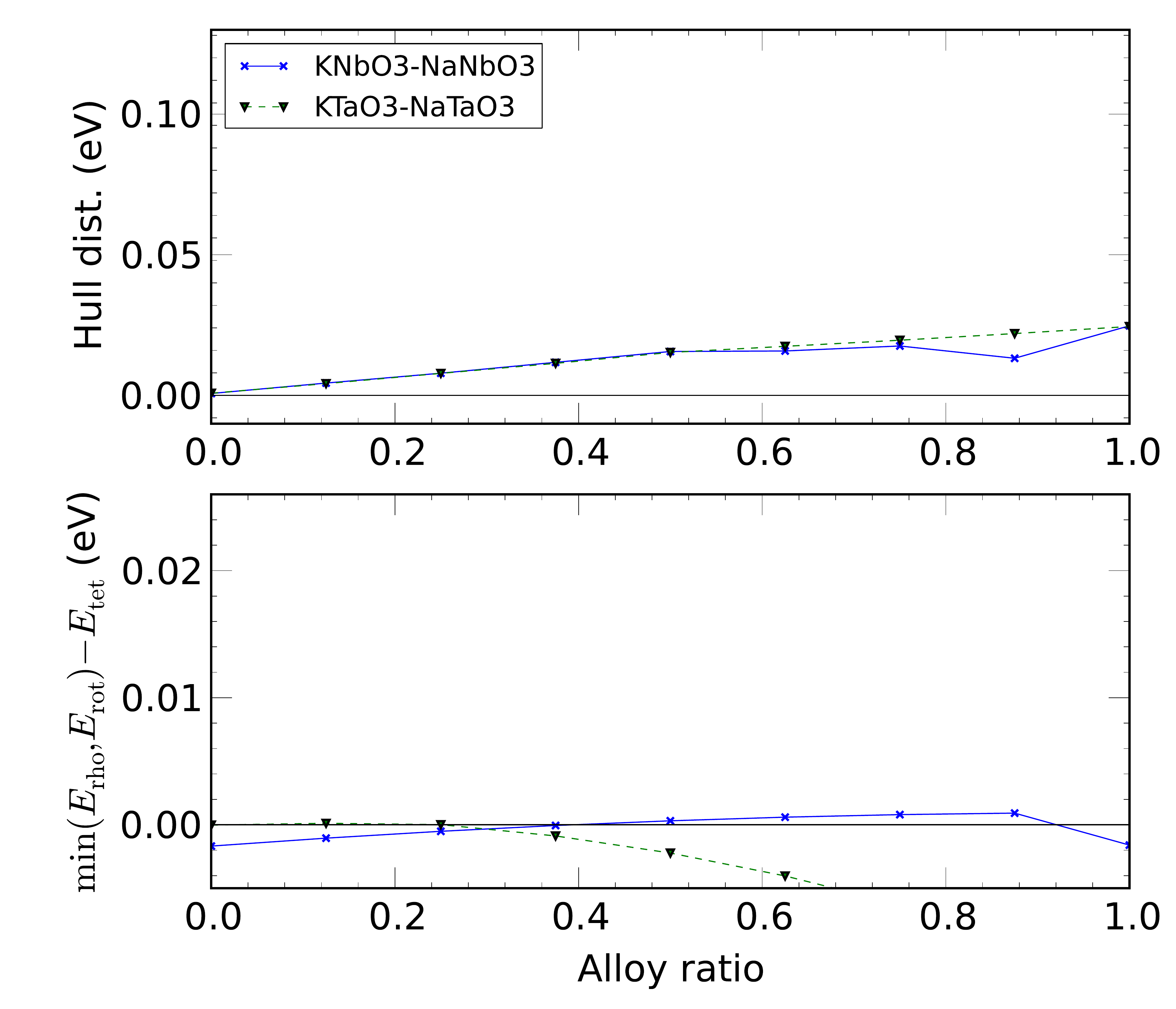}
\caption{(color online) Stability (upper graph) and energy preference for TET distortion (lower graph) in eV/atom for the $(\mathrm{K},\mathrm{Na})(\mathrm{Nb},\mathrm{Ta})\mathrm{O}_3$ system.}
\label{fig:toyota1}
\end{center}
\end{figure}

\begin{figure}
\begin{center}
\includegraphics[width=246.0pt]{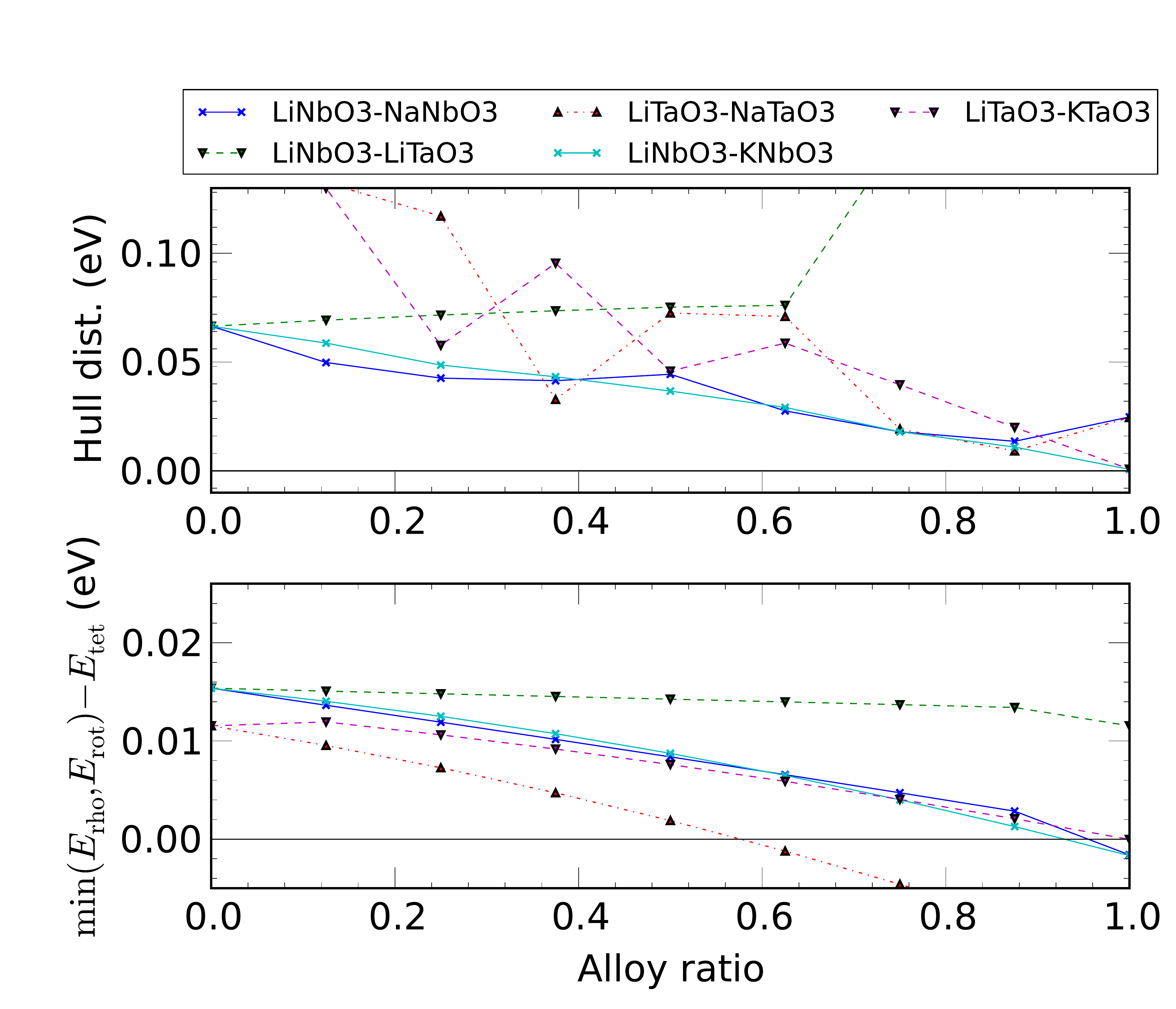}
\caption{(color online) Stability (upper graph) and energy preference for TET distortion (lower graph) in eV/atom for relevant alloys with Li with the components of the $\mathrm{(K,Na)(Nb,Ta)O}_3$ system. The system becomes less stable the more $\mathrm{Li}$ is added, but it also more strongly prefer the TET distortion.}
\label{fig:toyota2}
\end{center}
\end{figure}

Another known alloy system with an MPB has been identified by Saito {\it et al.}, $(\mathrm{K},\mathrm{Na},\mathrm{Li})(\mathrm{Nb},\mathrm{Ta})\mathrm{O}_{3}$ \cite{toyota}. Figure~\ref{fig:toyota1} and \ref{fig:toyota2} show our stability and distorted phase results for various two-component alloy systems related to this material. The base component is the $\mathrm{(K,Na)(Nb,Ta)O}_3$ system. According to the hull distance graph of Fig.~\ref{fig:toyota1} this alloy has a very small distance to the hull for all mixing ratios. The lower graph shows that the TET distortion and the closest competing distortion essentially share the same energy. In Fig.~\ref{fig:toyota2} we see various combinations of one of these end-points and $\mathrm{Li}$. The lower graph of Fig.~\ref{fig:toyota2} suggests a strong trend between $\mathrm{Li}$ concentration and a stronger preference for TET distortion \cite{singhprl}. However, in the hull distance graph, we see that higher concentration of $\mathrm{Li}$ increases the hull distance. 

From these results we confirm that synthesis of the material found by Saito {\it et al.} needs to strike a balance between $\mathrm{Li}$, which lowers the energy of a tetragonal distortion, with other components that make the alloy thermodynamically stable. 

\subsection{Other Pb-based alloys}
\begin{figure}
\begin{center}
\includegraphics[width=246.0pt]{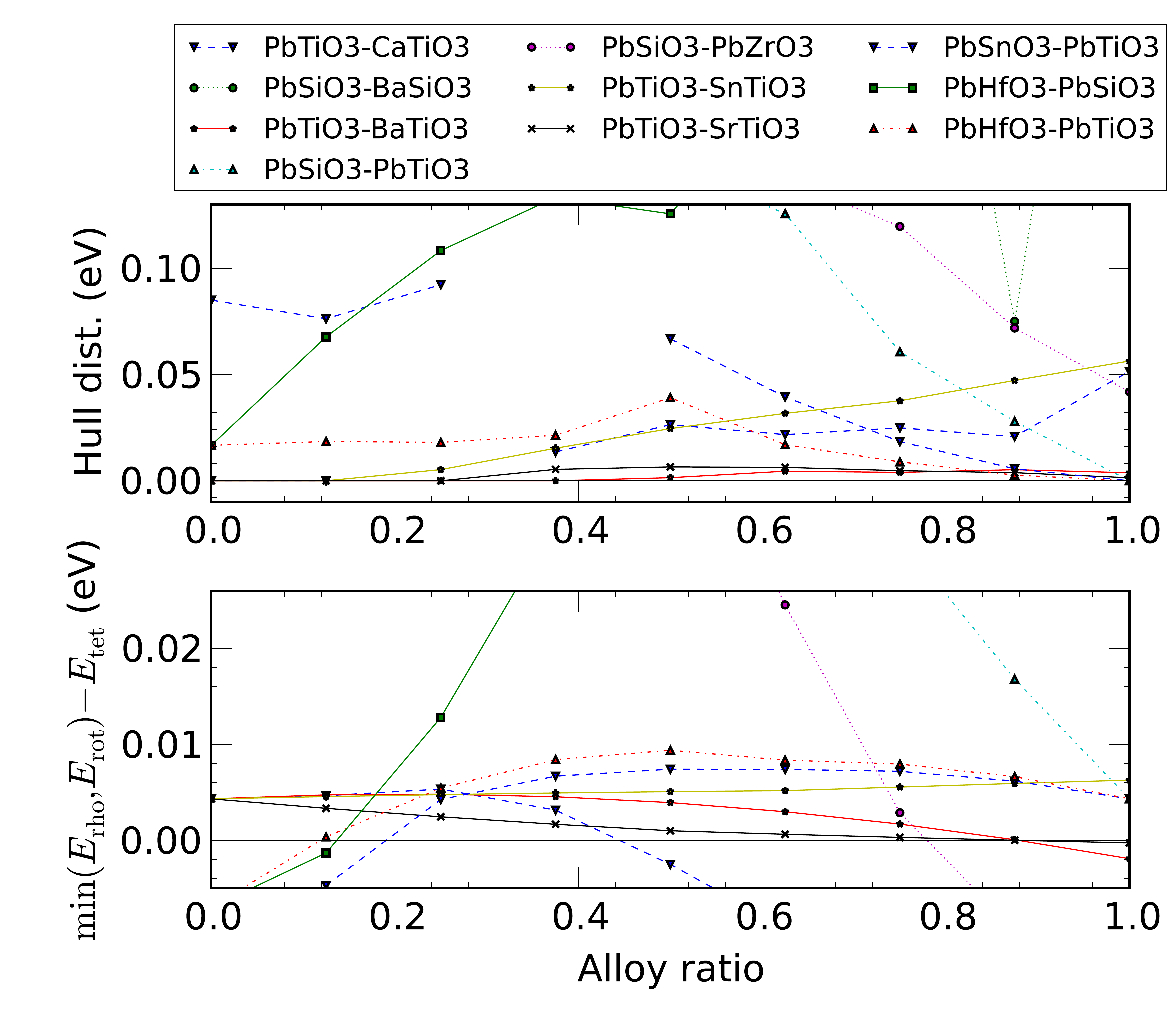}
\caption{(color online) Stability (upper graph) and energy preference for TET distortion (lower graph)  in eV/atom for all two-component alloys that remains after our initial screening which involve $\mathrm{Pb}$ (except for PZT, which is shown in Fig.~\ref{fig:pzt}). While individual alloys are not clearly discernible in the figure, it shows the general trend of these alloys as having small hull distances and TET distortions. For details on individual alloys, see the supplemental material to this paper \cite{supplemental}.}
\label{fig:pb}
\end{center}
\end{figure}

We have not excluded Pb-based alloys in our screening, and thus various combinations that include Pb show up as giving stable alloys that prefer the TET distortion. For the pursuit of lead-free piezoelectrics these may be of less interest, though, they are possibly relevant for reducing the relative amount of lead in such materials. An overview of the compounds considered in this work are presented in Fig.~\ref{fig:pb}. The figure shows many alloys with Pb that prefer the TET distortion, as seen by most lines in the lower graph being above zero, while simultaneously having comparably small hull distances, i.e., many of the corresponding lines in the hull distance graph are far below $0.5\ \mathrm{eV}$.

The results suggest it is common for alloys with Pb to form thermodynamically stable alloys that favors TET distortion. While this is not a surprising conclusion for perovskite oxides, it further adds to the confidence in the presented methods, and suggests our scheme may be able to reliably identify such general trends also in less studied perovskite families, e.g., nitrides.

\subsection{Cu-based alloys}
\begin{figure}
\begin{center}
\includegraphics[width=246.0pt]{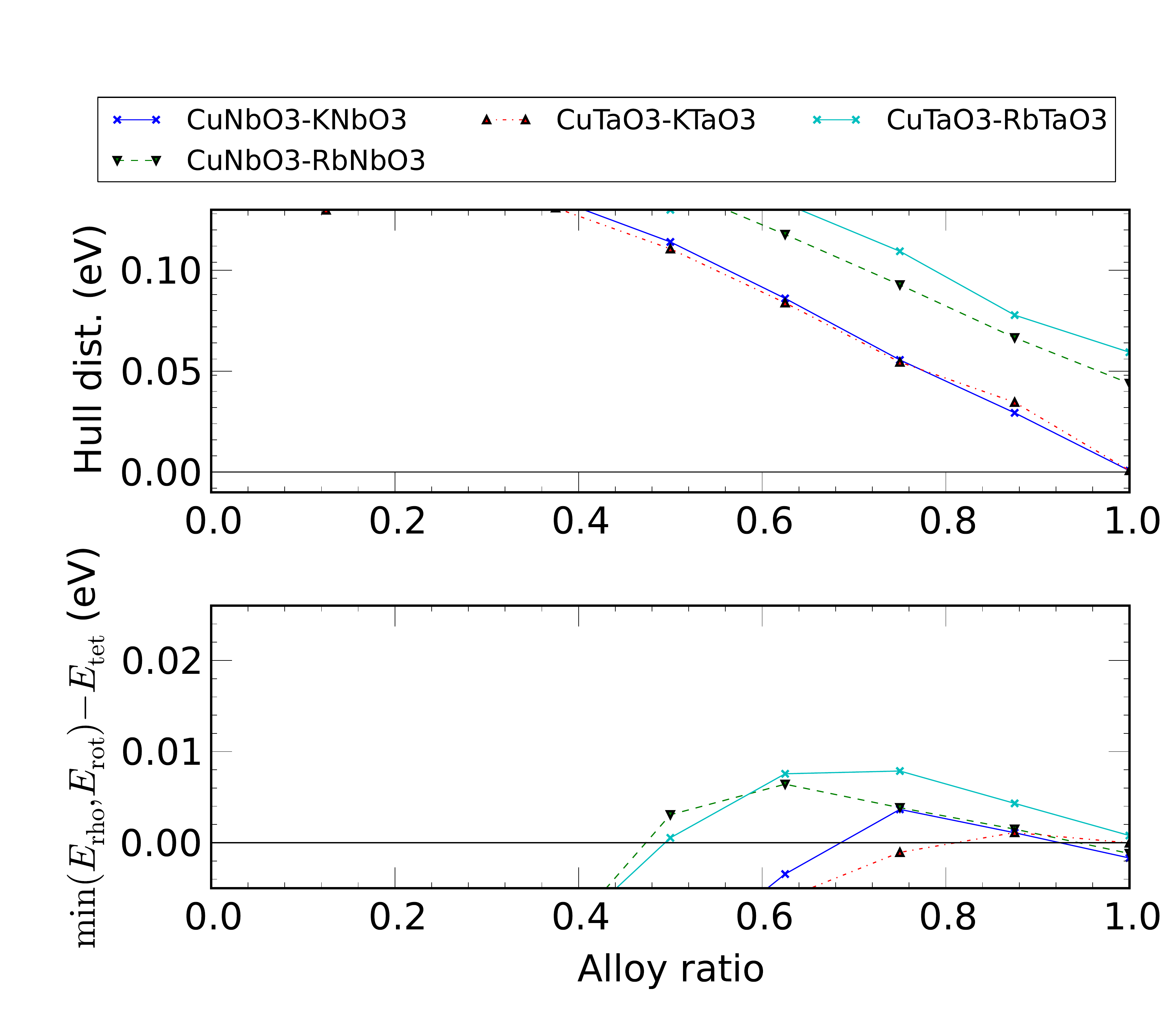}
\caption{(color online)  Stability (upper graph) and energy preference for TET distortion (lower graph) in eV/atom when $\mathrm{Cu}$ is introduced into various perovskites. Small to moderate amounts of $\mathrm{Cu}$ appear to make the alloys more strongly prefer the TET distortion, and there may be a region of such substitution into $\mathrm{K(Nb,Ta)O}_3$ with possibly formable alloys.}
\label{fig:other1b}
\end{center}
\end{figure}

Figure~\ref{fig:other1b} shows perovskite solid solutions with $\mathrm{Cu}$. The hull distance graph shows that low concentrations of $\mathrm{Cu}$ give hull distances below $0.5\ \mathrm{eV}$. The graph over energetics shows that it is also only at low concentrations of $\mathrm{Cu}$ such alloys prefer the TET distortion.

Hence, it appears small amounts of $\mathrm{Cu}$ on the A-site pushes the perovskite structure towards a TET distortion, but at high amounts the TET distortion becomes energetically unfavored again. Especially in $\mathrm{K(Nb,Ta)O}_3$ it appears to be an interval of $\mathrm{Cu}$ mixing ratio that does not substantially increase the distance to the hull, i.e., where $\mathrm{Cu}$ behaves similar to $\mathrm{Li}$ in the KNLNT system. These results suggest that it may be interesting to further investigate the $(\mathrm{Cu}_x\mathrm{K}_{1-x})(\mathrm{Nb},\mathrm{Ta})\mathrm{O}_3$ alloy system for small values of $x$. 

\subsection{Si-based alloys}
\begin{figure}
\begin{center}
\includegraphics[width=246.0pt]{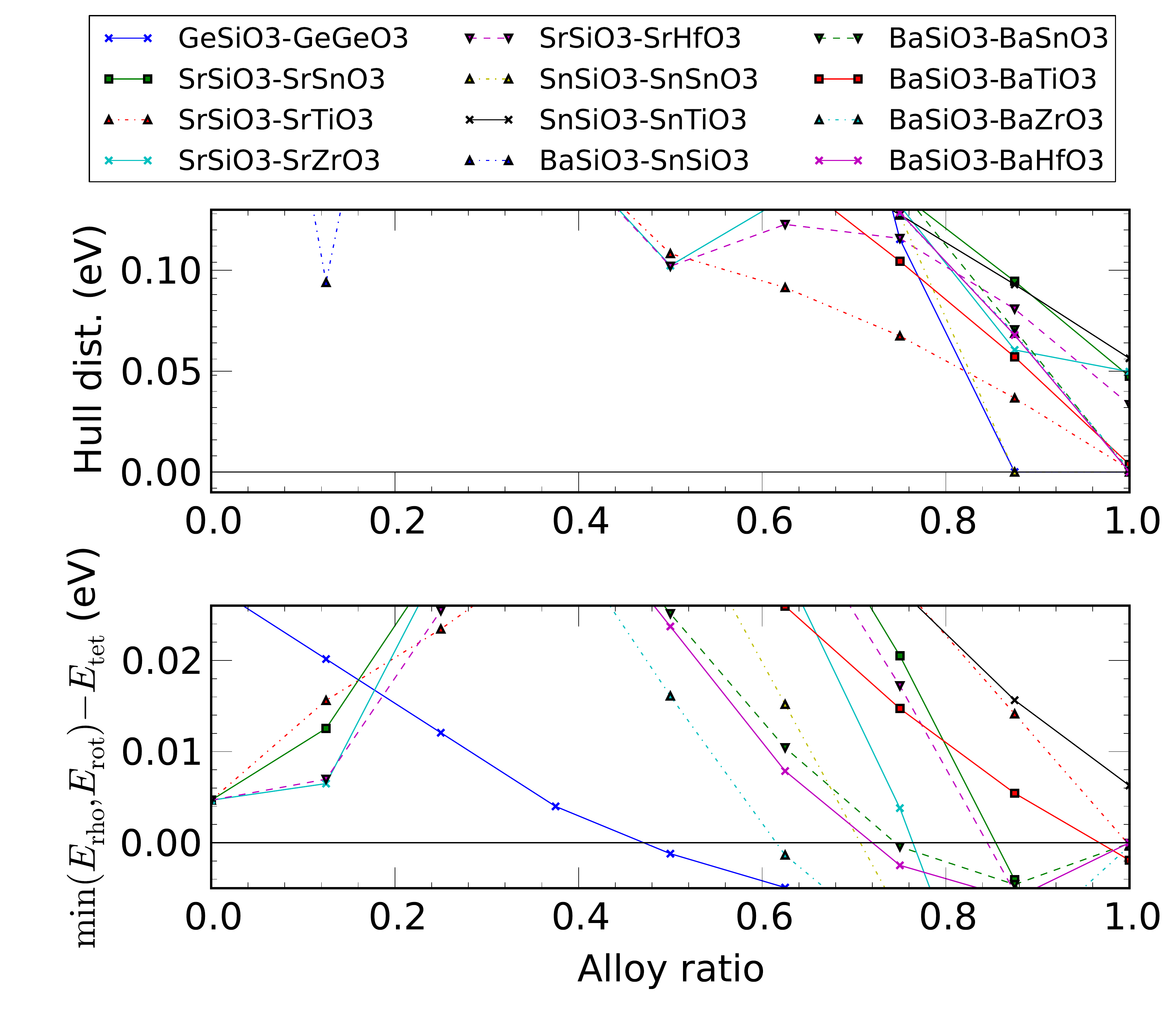}
\caption{(color online) Stability (upper graph) and energy preference for TET distortion (lower graph)  in eV/atom for two-component alloys involving $\mathrm{Si}$. The $\mathrm{Si}$-based end-points are generally unstable. While individual alloys are not clearly discernible in the figure, the upper graph shows a general trend of these end-points to not mix well. For details on individual alloys, see the supplemental material to this paper \cite{supplemental}.}
\label{fig:other2}
\end{center}
\end{figure}

According to our previous investigation, perovskites with $\mathrm{Si}$ or $\mathrm{Ge}$ on the B-site, and in particular $\mathrm{BaSiO}_3$, were found to have promising properties as alloy end-points \cite{armientoprb}. The lower part of Fig.~\ref{fig:other2} shows that the majority of the $\mathrm{Si}$-based alloys studied in this work have moderate to strong preference for a TET distortion. The upper graph of Fig.~\ref{fig:other2} shows that both the end-points and alloys have large hull distances. 

Hence, we confirm that when the preference for TET distortion is considered, $\mathrm{Si}$-based compounds look interesting. However, it appears these systems are generally not thermodynamically stable in TET distortion.

\begin{figure}
\begin{center}
\includegraphics[width=246.0pt]{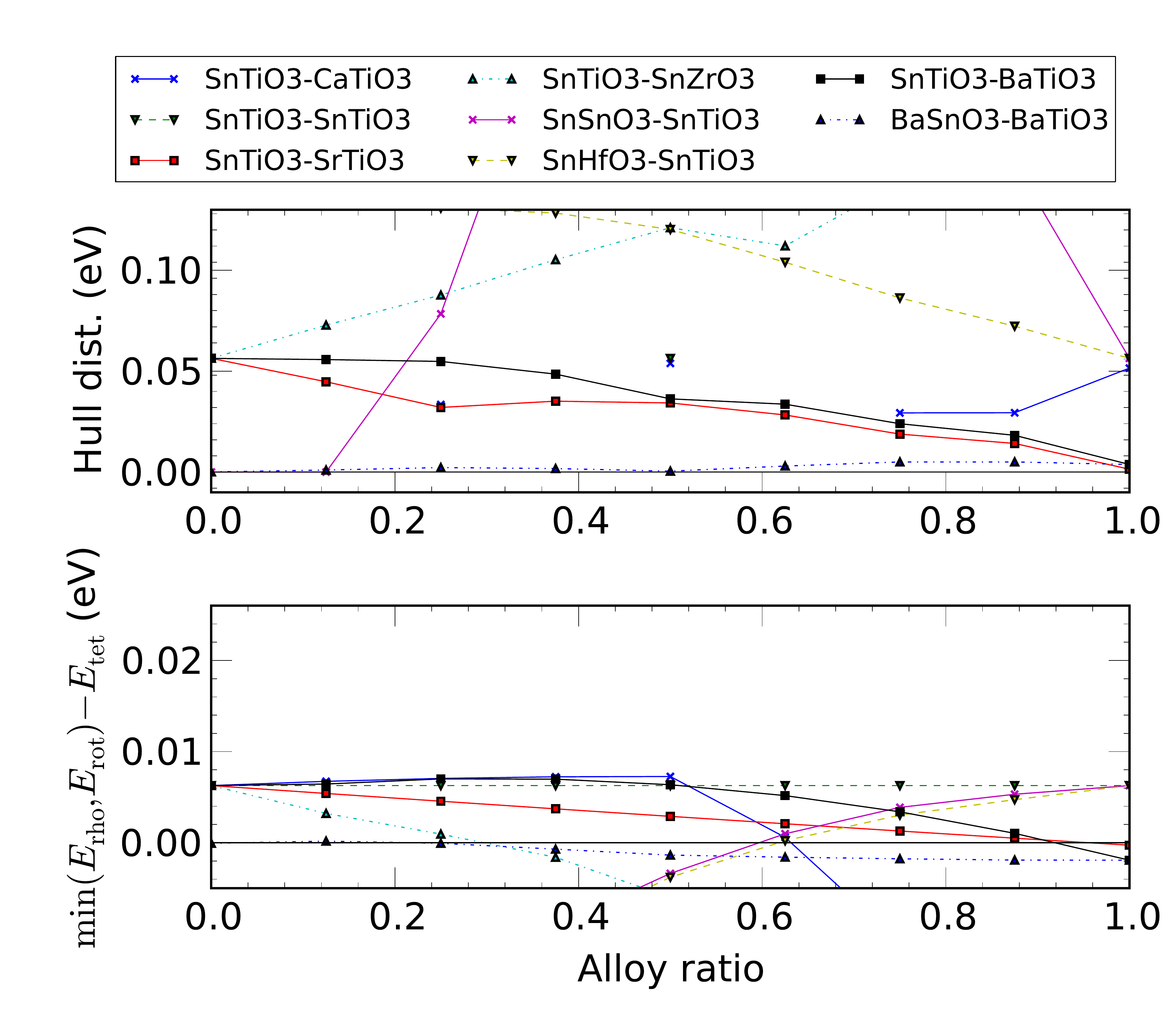}
\caption{(color online) Stability (upper graph) and energy preference for TET distortion (lower graph)  in eV/atom for two-component alloys involving $\mathrm{Sn}$. The alloy $\mathrm{(Ba_xSn_{x-1})TiO}_3$ stands out as a tetragonal alloy of promising stability, indicating that synthesis may be possible. $\mathrm{Ba(Sn_{x}Ti_{x-1})O}_3$ is even more stable and will generally be a readily accessible phase in a synthesis attempt of the former alloy.}
\label{fig:other4}
\end{center}
\end{figure}

\begin{figure}
\begin{center}
\includegraphics[width=246.0pt]{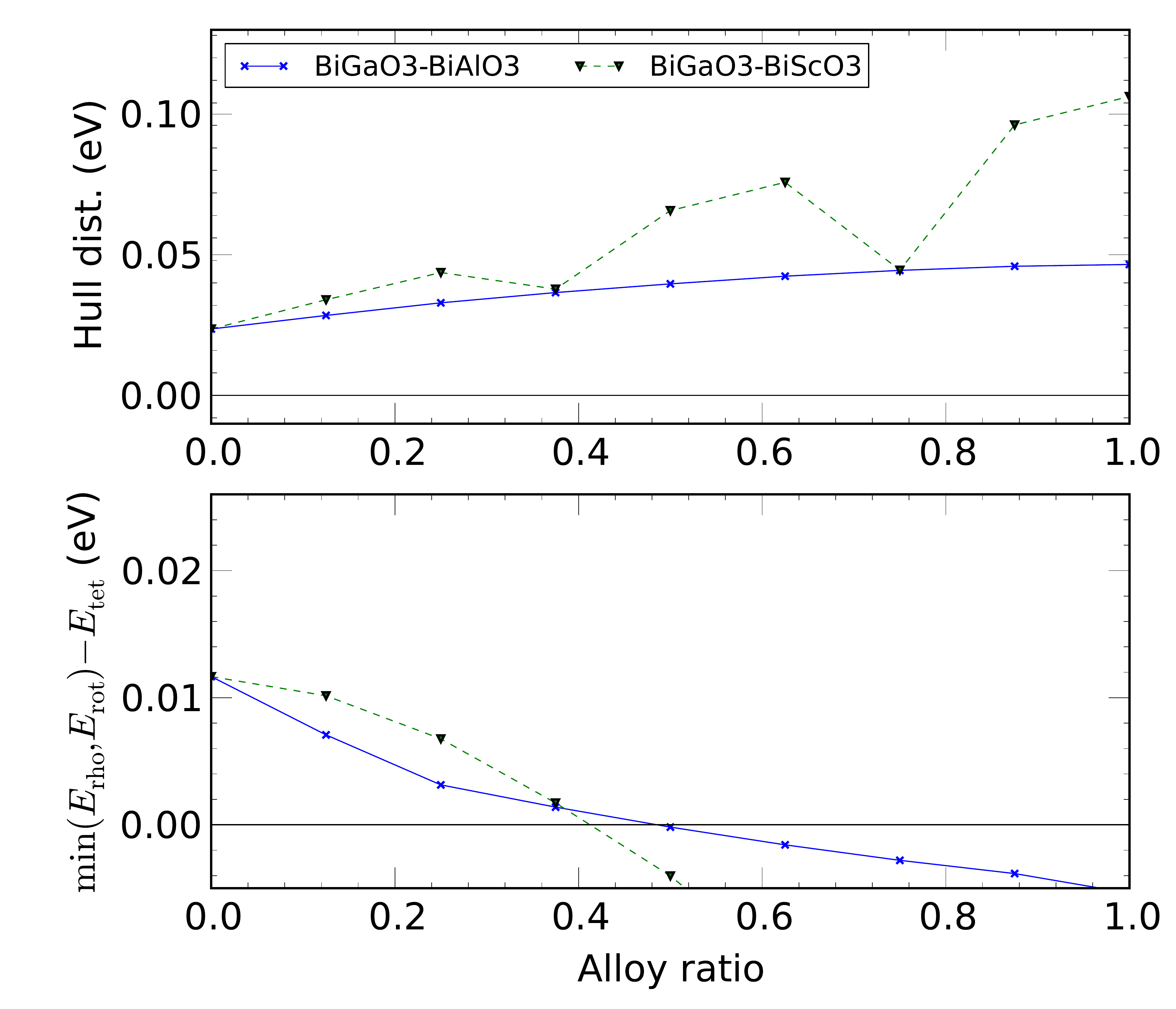}
\caption{(color online) Stability (upper graph) and energy preference for TET distortion (lower graph)  in eV/atom for two-component alloys involving $\mathrm{Bi}$.}
\label{fig:other5}
\end{center}
\end{figure}

\subsection{Sn-based alloys}
The possible synthesis of ferroelectric $\mathrm{Sn}$-based alloys is an active area of research (see, e.g., Refs.~\onlinecite{sn, sn2, bsite, gexp}). Figure~\ref{fig:other4} shows our results for $\textrm{Sn}$ based compositions. We note that in general, these compositions behave similar to the $\textrm{Pb}$-based alloys discussed above. The lower graph show many systems as preferring TET distortion, and the hull distance graph suggests that several have comparably small hull distances. In particular, we find small hull distances for the following alloys, $(\mathrm{Ba}_x\mathrm{Sn}_{1-x})\mathrm{Ti}\mathrm{O}_3$, $(\mathrm{Sn}_x\mathrm{Sr}_{1-x})\mathrm{Ti}\mathrm{O}_3$ and $\mathrm{Ba}(\mathrm{Sn}_x\mathrm{Ti}_{1-x})\mathrm{O}_3$. The last one has $\textrm{Sn}$ placed on the B-site rather than the A-site, and a significantly smaller hull distance than the other compositions.

These results confirm the interest in $\mathrm{Sn}$-based systems. We identify as promising especially the $(\mathrm{Ba}_x\mathrm{Sn}_{1-x})\mathrm{Ti}\mathrm{O}_3$ alloy with a large $x$ and possibly also $(\mathrm{Ca}_x\mathrm{Sn})\mathrm{Ti}\mathrm{O}_3$ with $x \sim 0.2$. However, for the former, we note that the lower hull distance of the $\mathrm{BaSnO}_3$ indicates that in solid solutions the $\mathrm{Sn}$ cations may tend to go for the $\mathrm{B}$-site in the synthesis. On the other hand, it may be interesting to study the stable $\mathrm{Ba(Sn_xTi_{1-x})O}_3$ system in more detail. In the interval $x\approx[0.05,0.2]$ our results predict this alloy to (very weakly) prefer a TET distortion. Furthermore, the $\mathrm{Ba(Hf_xTi_{1-x})O}_3$ system behaves very similarly (not shown here, see the supplemental material \cite{supplemental}), and hence these alloys may be useful in attempts at tuning the BTZC system discussed above.

\subsection{Bi-based alloys}
Finally, we turn to perovskite alloys with $\mathrm{Bi}$ on the $A$-site. The ones that remain after screening are $\textrm{Bi}(\textrm{Ga}_x\textrm{Al}_{1-x})\textrm{O}_3$ and $\textrm{Bi}(\textrm{Ga}_x\textrm{Sc}_{1-x})\textrm{O}_3$. The upper graph of Fig.~\ref{fig:other5} suggests both these alloys have fairly large hull distances. The lower graph shows that the TET distortion is preferred if the ratio of $\textrm{Ga}$ is large, i.e., $(1-x)<0.5$. 

The preference for TET distortion for both of these alloys has a composition dependence that resembles that of PZT. These systems may be of interest if they are possible to synthesize.

\section{Discussion}
The focus of this work is on finding a thermodynamically stable ferroelectric perovskite alloy that prefers a TET distortion. It is, perhaps, unsurprising that we do not identify any candidates equally promising as the ones already known in the well-studied family of perovskite oxides. Apart from the individual discussion presented above for each specific group of perovskite alloys, we make below a few general observations about our results and their generality.

The strategy behind the material by Saito {\it et al.} \cite{toyota}, $(\mathrm{K},\mathrm{Na},\mathrm{Li})(\mathrm{Nb},\mathrm{Ta})\mathrm{O}_{3}$, can be summarized as starting from a thermodynamically stable composition that does \emph{not} prefer the TET distortion, $\mathrm{K}(\mathrm{Nb},\mathrm{Ta})\mathrm{O}_{3}$. This system is mixed with $(\mathrm{Li})(\mathrm{Nb},\mathrm{Ta})\mathrm{O}_{3}$ which prefer the TET distortion, but is thermodynamically unstable on perovskite form. Hence, larger amounts of $\mathrm{Li}$ makes the alloy more strongly prefer the TET distortion, but it also makes it less stable. This mechanism is directly seen in Fig.~\ref{fig:toyota2}, where an increased concentration of $\textrm{Li}$ correlate both with a higher energy preference for a TET distortion and an increased hull distance. We identified that a similar mechanism may be present when a small amount of $\mathrm{Cu}$ is introduced in $\mathrm{K(Ta,Nb)O}_3$ as seen in Fig.~\ref{fig:other1b}. Since this mechanism essentially only relies on finding an unstable component that strongly prefers the TET distortion, it appears likely to find it also in systems not yet studied with our formalism, e.g., (oxy-)nitride perovskites, perovskites with rare earth elements, and, possibly, other crystal structures that can accommodate a MPB.

On the other hand, the TET distorted material used in the work by Liu and Ren \cite{prl}, $(\mathrm{Ba}_{x}\mathrm{Ca}_{1-x})\mathrm{Ti}\mathrm{O}_3$ shows a less clear relation between concentration, energy preference for the TET distortion, and hull distance. Here, two thermodynamically stable materials are mixed, neither which prefer the TET distortion. The component with the smaller cell volume will make the alloy prefer the TET distortion (see, e.g., Ref.~\onlinecite{armientoprb} for an expanded discussion.) This is a very attractive mechanism for creating a TET distorted alloy since it avoids involving an unstable end-point in the alloy and thus seems likely to be easier to synthesize. However, we have not identified this mechanism in any other of the studied 2-component perovskite oxides (possibly excluding some $\textrm{Pb}$-based systems; note how in Fig.~\ref{fig:pzt} PZT is predicted to have a larger preference for the TET distortion than PT itself.) It remains to be seen if the mechanism of the Liu and Ren material is more common in any of the other systems not yet studied with our methodology.

\section{Summary and Conclusions}

In this work we have used a large-scale high-throughput DFT framework to investigate an extensive space of perovskite alloys. Our investigation has focused on identifying two-component alloys with a ground state TET distortion, i.e., candidates for the primary component in a system with an MPB that most prominently combine a preference towards TET distortion with stability are alloys which have already been used or proposed for high-performance piezoelectrics. However, we identify some promising less studied systems. In particular, $\mathrm{K(Ta,Nb)O}_3$ with a small amount of $\mathrm{Cu}$; and $\mathrm{BiGaO}_3$ with small amounts of $\mathrm{Sc}$ or $\mathrm{Al}$. We also confirm the ongoing research interest into $\mathrm{(Ba,Sn)}$-based titanates to be highly relevant.

The presented interpolation scheme successfully describes the energetics of distorted phases of isovalent alloys with enough accuracy to identify materials with a specific ground state distortion. Our results point at the space of oxide perovskite as mostly exhausted for novel MPB systems. The formalism appear promising to extend into into less experimentally explored perovskite systems, e.g., (oxy-)nitrides, perovskites with rare-earth elements, or even to identify MPBs in other crystal structures.

In high-throughput material design, it is a major challenge to handle alloy systems, since they can quickly lead to very extensive chemical spaces. The method presented in this paper demonstrates a way to predict the energetics of distorted phases of isovalent alloys that avoids extensive brute-force computational work that would be prohibitively time consuming. The discussed approach may also be very useful also for high-throughput material design in other areas.

\begin{acknowledgments}
The authors acknowledge funds provided by Robert Bosch LLC. R.A. also acknowledge funds provided by the Swedish Research Council grant 621-2011-4249 and the Linnaeus Environment at Link{\"o}ping on Nanoscale Functional Materials (LiLi-NFM) funded by VR. G.H. acknowledges financial support from FRS-FNRS.
\end{acknowledgments}

\end{document}